
\input phyzzx
\nopubblock
\sequentialequations
\twelvepoint
\overfullrule=0pt
\tolerance=5000

\REF\hawking{S. W. Hawking. Commun. Math. Phys. {\bf 43} (1975), 199.}

\REF\birrel{See e.g. N. D. Birrell and P. C. W. Davies, {\it Quantum Fields
in Curved Space} (Cambridge University Press, Cambridge, 1982).}

\REF\israel{See W. Israel, in {\it 300 Years of Gravitation\/} (Cambridge
University Press, Cambridge, 1987), especially page 234.
The line element first appears in work of Painleve and Gullstrand in 1921.
We thank Ted Jacobson for bringing this reference to our attention.}

\REF\bcmn{B. K. Berger, D. M. Chitre, V. E. Moncrief, and Y. Nutku. Phys. Rev.
{\bf D5} (1972), 2467.}

\REF\polch{W. Fischler, D. Morgan, and J. Polchinski. Phys. Rev. {\bf D42}
(1990), 4042.}

\REF\unruh{W. G. Unruh. Phys. Rev. {\bf D14} (1976), 870.}

\REF\psstw{J. Preskill, P. Schwarz, A. Shapere, S. Trivedi, and F. Wilczek.
Mod. Phys. Lett. {\bf A6} (1991), 2353.}

\REF\affleck{I. K. Affleck, O. Alvarez, and N. S. Manton. Nucl. Phys.
{\bf B197} (1982), 509.}

\FIG\diag{Penrose diagram for the Schwarzschild geometry.  As described
in the text, $r$ and $t$ in one coordinate patch, for the upper sign in
the line element, cover regions I and II.  As is clear from the
diagram, the ingoing light rays are captured in their entirety (into the
singularity), whereas the outgoing light rays cannot be traced back
past the
horizon.}

\line{\hfill }
\line{\hfill PUPT 1474, IASSNS 94/46}
\line{\hfill gr-qc/9406042}
\line{\hfill June 1994}

\titlepage
\title{Some Applications of a Simple Stationary
Line Element for the Schwarzschild Geometry}

\author{Per Kraus\foot{~~~perkraus@puhep1.princeton.edu}}
\vskip .2cm

\centerline{{\it Department of Physics}}
\centerline{{\it Joseph Henry Laboratories }}
\centerline{{\it Princeton University}}
\centerline{{\it Princeton, N.J. 08544}}

\author{Frank Wilczek\foot{Research supported in part by DOE grant
DE-FG02-90ER40542.~~~WILCZEK@SNS.IAS.EDU}}
\vskip.2cm

\centerline{{\it School of Natural Sciences}}
\centerline{{\it Institute for Advanced Study}}
\centerline{{\it Olden Lane}}
\centerline{{\it Princeton, N.J. 08540}}
\endpage

\abstract{Guided by a Hamiltonian treatment
of spherically symmetric geometry, we are led to
a
remarkably simple -- stationary, but not
static -- form for the line element of
Schwarzschild (and Reissner-Nordstrom)
geometry.
The line element continues
smoothly through the horizon; by exploiting this feature we are able
to give a very simple and physically transparent derivation of the
Hawking radiance.  We construct
the complete Penrose diagram by enforcing time-reversal
symmetry.  Finally
we outline how an improved treatment of the radiance, including
effects of self-gravitation, can be obtained.}

\endpage

Schwarzschild found his remarkable exact solution for
the geometry outside a star in general relativity
quite soon after Einstein
derived the field equations.  Further
study of this geometry over the course of several decades
revealed a series of surprises: the existence and physical
relevance of pure vacuum ``black hole'' solutions; the
incompleteness of the space-time covered by the
original Schwarzschild coordinates, and the highly
non-trivial global structure of its completion; and the
dynamic nature of the physics in this geometry despite its
static mathematical form, revealed perhaps most dramatically
by
the Hawking radiance [\hawking ].  Discussions
of this material can now be found
in advanced
textbooks [\birrel ], but they are hardly limpid.

In the course of investigating an improvement to the standard
calculation of this radiance to take into account its
self-gravity,
as we shall sketch below,
we came upon a remarkably simple form for the line element of
Schwarzschild (and Reissner-Nordstrom) geometry.
This line element has an interesting history
[\israel ], but
as far as we know it has never been discussed from a modern
point of view.
We have found
that several of the more subtle features of the geometry become
especially easy
to see when this line element is used.

\chapter{Hamiltonian Form, and Derivation}

The general spherically symmetric line element can be written
in the form
$$
ds^2~=~-(N^t)^2 dt^2 +  L^2 (dr + N^r dt )^2 +
R^2 (d\theta^2  +  \sin^2 \theta d\phi^2 )~.
\eqn\lineelt
$$
In this expression it is to be understood that $L$, $R$,
$N^t$ and $N^r$ are functions of $r$ and $t$.  When one inserts
this form into the Einstein-Hilbert action, one finds that
no time derivatives of the lapse $N^t$ or of the shift $N^r$ occur,
so that variation of these quantities generates constraints.
These are [\bcmn ,\polch ]
$$
{\cal H}^G_t~=~L\pi_L^2/ 2R^2 - \pi_L \pi_R / R +
\bigl( {R R^\prime \over L } \bigr)^\prime
- \bigl( {R^{\prime 2 }\over 2L }\bigr) - L/2 ~=~ 0
\eqn\lapseconstraint
$$
and
$$
{\cal H}^G_r~=~R^\prime \pi_R - L \pi_L^\prime ~=~ 0~,
\eqn\shiftconstraint
$$
where $\prime $ represents ${d\over dr}$ and
$\dot {}$ represents ${d\over dt}$, and of course
$\pi_L = (N^r R^\prime - RR^\cdot )/ N^t$,
$\pi_R = (N^rLR)^\prime/N^t - (LR)^\cdot /N^t$
are the canonical conjugates of $L$, $R$
respectively.

Furthermore the action is invariant under reparametrization,
so that one should put extra restrictions on
$L$ and $R$ (fix a gauge) in order to eliminate
spurious degrees of freedom.  The form of the solutions
obtained, though
not their physical content, will depend on one's choice
for these
restrictions.

Our choice is simply $L=1$, $R=r$.  With this choice the
equations simplify drastically, and
one easily solves to find
$\pi_L = \sqrt {2Mr }~,~ \pi_R = \sqrt {M\over 2r}$
and then $N^t = \pm 1,~ N^r = \pm \sqrt {2M\over r}$.
Thus for the line element we have
$$
ds^2~=~ -dt^2 + (dr \pm \sqrt {2M \over r} dt )^2 +
r^2 (d\theta^2  +  \sin^2 \theta d\phi^2 )~.
\eqn\nicemetric
$$
$M$, which appears as an integration constant, of course
is to be interpreted as the mass of the black hole described by
this line element.

For the Reissner-Nordstrom geometry, the same gauge choice leads
to a metric of the same form, with the only change that
$2M \rightarrow 2M - Q^2/r$.

These line elements are stationary -- that is, invariant under
translation of $t$,
but not static -- that is, invariant under reversal
of the sign of
$t$.  Indeed reversal of this sign interchanges the $\pm$
in \nicemetric , a feature we will interpret further below.
Another peculiar feature is that each
constant time slice
$dt=0$ is simply flat Euclidean space!

\chapter{Global Structure}


Now let us discuss the global properties of our coordinate system.
Perhaps the clearest approach to such questions is via
consideration of the properties
of light rays.  Taking for definiteness the upper sign in \nicemetric ,
and without any essential
loss of generality restricting to the case
$d\theta = d\phi =0$ appropriate to the $\theta = \pi/2$ sections,
we find that $ds^2=0$ when
$$
{dr\over dt } ~=~ - \sqrt {2M\over r } \mp 1~.
\eqn\lightrays
$$
For the class of light rays governed by
the upper sign, we can cover the
entire range $0 < r < \infty $ as $t$ varies.  In particular one
meets
no obstruction, nor any special structure,
at the horizon $r=2M$.  For the class of light rays
governed by the lower sign there is structure at $r=2M$.
When $r > 2M$ one has a positive slope for ${dr\over dt}$, and
$r$ ranges over $2M < r < \infty$.  When $r < 2M$ one has a negative
slope for ${dr\over dt}$, and $r$ ranges over $0 < r < 2M$.
When $r = 2M$ it does not vary with $t$.  From these properties, one
infers that our light rays cover regions I and II in the Penrose
diagram, as displayed in Figure 1.  Let us emphasize that the
properties of the Penrose diagram can be {\it inferred\/} from the
properties of the light rays, although we will not belabor that
point here.


If one chooses instead the lower sign in \nicemetric , and performs
a similar analysis, one finds that regions I and II$^\prime$
are covered.
Patching these together with the sectors found previously,
one still does not have a complete space-time.
However our line element is not yet exhausted.  For in drawing
Figure 1 we have implicitly assumed that $t$ increases along light
rays which point up (``towards the future'').  Logically, and to
maintain symmetry, one should
consider also the opposite case, that the coordinate
$t$ increases towards the past.  By doing this, one generates
coordinate systems covering regions
I$^\prime$ and II$^\prime$
respectively I$^\prime$ and II$^\prime$,
for the upper and lower signs
in \nicemetric .   Thus the complete Penrose diagram is covered
with patches each governed by a stationary -- but not static --
metric, and with non-trivial regions of overlap.


In the Reissner-Nordstrom case the generalization of \nicemetric\
has
a coordinate singularity at $r=Q^2/2M$.  However this singularity
is inside both horizons, and does not pose a serious obstruction to
a global analysis.  One obtains the complete Penrose diagram also in
this case by
iterating
constructions similar to those just sketched.


The usual
Schwarzschild line element appears to be time reversal symmetric,
but when the global structure of the space-time it defines is taken into
account one sees that this appearance is misleading.  The fully extended
light-rays in Figure 1 go from empty space to a singularity as $t$
advances (they pass from region I into region II),
which is definitely distinguishable from the
reverse process.  There is a
symmetry which relates these to the corresponding rays going from
region II$^\prime$ to region I$^\prime$, however it involves not merely
changing the sign of
$t$ in the Schwarzschild metric,
but rather going to a completely disjoint
region of the space-time.  This actual symmetry of the space-time is if
anything more obvious in our construction than in the standard one.
Thus by taking the line-element
in region I stationary rather than static
we have lost some false symmetry while making the true symmetry --
and its
necessary
connection with the existence of
region I$^\prime$ (constructed, as we have seen, by simultaneously
reversing the
sign of $t$ {\it and\/} interchanging the future with the past) --
more obvious.

\chapter{Boundary Conditions and Radiance}

Taking with the upper sign in \nicemetric\ and the normal
time-orientation,
we have a coordinate system that goes through the horizon
smoothly
and contains future infinity.  We can
use it to
discuss the problem of
defining boundary conditions on the quantum
fields, such that a freely falling observer will see
no singular
behavior when passing through the horizon.
This presumably corresponds to the physical situation
for the geometry defined by collapse, since there is nothing
singular or special in the {\it local\/} geometry at the horizon
-- and indeed, strictly speaking
the position of the horizon actually depends on
future events!

Thus we seek to construct a vacuum state which has a
non-singular stress-energy as measured by freely falling
observers.  For concreteness, let us consider how we should
construct the vacuum state for
a massless scalar field $\phi$.
By the equivalence principle, and standard quantum field
theory in flat space, we should leave all the positive-frequency
modes empty -- where positive frequency is defined in a coordinate
system that is locally flat.  For the metric we are considering,
it is convenient to work along a curve
$dr+ \sqrt {2M\over r} dt =0$; then the condition is simply positive
frequency with respect to $t$ near this curve.

Now at spatial infinity (more accurately: conformal infinity
${\cal I}_+$) the vacuum state is defined
locally by the requirement that
modes having positive frequency with respect to the variable
$u = t_{\rm s} -r_*$ are unoccupied, where $t_{\rm s}$
is Schwarzschild
time and $r_* = r + 2M\ln (r -2M)$ is the tortoise coordinate.
We wish to find the relationship between this
requirement and the preceding one.

The relationship between $t$ and $t_{\rm s}$ is
$$
t = t_{\rm s} + 2 \sqrt{2Mr} +
  2M \ln {{\sqrt r - \sqrt{2M} } \over {\sqrt r + \sqrt{2M} }}
\eqn\times
$$
so that
$$
u = t_{\rm s} - r_* =
t - 2 \sqrt{2Mr} -r -4M \ln (\sqrt r - \sqrt{2M} )~,
\eqn\uexpression
$$
and thus one finds that along a curve with $dr+ \sqrt {2M\over r} dt =0$,
$$
{du\over dt}~=~ 2 + \sqrt {2M\over r} + {2M \over r - \sqrt{2Mr} }~.
\eqn\frequencyfactor
$$
Because the last term on the right hand side
is singular, the two definitions of positive
frequency -- with respect to $u$ or to $t$ -- do not coincide.
To remove the singularity, note that along any of
the curves of interest
$e^{-u/4M}$ has a simple zero at $r=2M$, but is otherwise positive.
Clearly then demanding positive frequency with respect to
$t$ along such curves requires positive frequency not with respect
to $u$ but rather with respect to
$$
U = - e^{-u/4M}~.
\eqn\kruskalU
$$
In this way we have arrived at the famous Unruh
boundary conditions [\unruh ]. From these boundary conditions, one readily
derives the Hawking radiance.

\chapter{Radiance: Dynamical Treatment}

In the previous section, we have discussed the implementation of
physical boundary conditions on external quantum fields, regarding
the background geometry as fixed and given.  For many reasons, one
would like to go beyond this treatment, and to consider the geometry
also as a quantum variable.   In its full generality this problem
seems out of reach at present.   However there is a
situation in which the standard treatment breaks down for
a simple concrete physical reason, which
plausibly can be repaired by a relatively
modest expansion of the formalism.

We are referring to the problem of
radiance of a near-extremal Reissner-Nordstrom hole [\psstw ].
As one approaches extremality ($Q \rightarrow M$),
the formal temperature $T$ of the Hawking radiation approaches
zero, in such a way that radiation of a single typical
quantum of energy $T$ changes the temperature by an amount large
compared to its value.  In such circumstances, the approximation
of treating the geometry as fixed is obviously inadequate, since
it matters very much in the formulae whether $M$ is the mass
before or after the radiation of a quantum.
This ambiguity will be removed if one expands the formalism to
include the effects of the gravitational self-interaction of
the radiation in a consistent way.

With this motivation, and also simply to have a
tractable but non-trivial problem in quantum geometry,
we consider the following truncation of the full dynamical
problem.  We model the s-wave dynamics of an emitted particle by quantizing a
spherically symmetric self-gravitating shell around the black hole.  The
classical action for this system is the sum of the action for the shell itself,
and the action of the gravitional field it produces.  To quantize, we have
found it most convenient to use a Hamiltonian path integral description, for
which the action takes the form
$$
S = \int dt~ p_{\hat{r}}~ \dot{\hat{r}}~+~\int dt~dr~[\pi_{L}\dot{L} +
\pi_{R}\dot{R} - N^t ({\cal H}^G_t + {\cal H}^M_t) - N^r({\cal H}^G_r+
{\cal H}^M_r)] ~-~\int dt~ M
\eqn\action
$$
where
$$
{\cal H}^M_t = \sqrt{{{p_{\hat{r}}}^2 \over L^2} + {m_o}^2}~\delta(r-\hat{r}),
$$
$$
{\cal H}^M_r = - p_{\hat{r}}~\delta(r-\hat{r}),
\eqn\Hdef
$$
and ${\cal H}^G_t$, ${\cal H}^G_r$ are given by \lapseconstraint\  and
\shiftconstraint.
In the above, $\hat{r}$ is the coordinate of the shell, $m_o$ its rest mass,
and $M$ is the ADM mass of the system.  Once again, varying with respect to
$N^t$ and $N^r$ yields constraints, but now the constraints depend on the
shell variables as well. To obtain an effective action depending only on the
position and momentum of the shell, we can solve the constraints to find
expressions for the metric variables in terms of the shell variables, and then
insert these relations back into the action.  Quantization of this action
leads to a wave equation modified to include self-interaction, as will be
detailed in a forthcoming publication.

   The physical rationale for considering the above model is that for a large
($M>>M_{\rm Pl.}$) near extremal hole the only important fluctuations in the
geometry come from the occasional emission of a quantum.  It then seems
reasonable to treat the collapsing matter that formed the hole classically.
Furthermore, since the hole is cold the radiation is sporadic, so the
approximation of including self-interaction
while ignoring interparticle interactions would appear to be a good
one. In this way we arrive at a picture reminiscent of the path integral
treatment of pair production at strong coupling in a weak electric field
[\affleck ], where an effective action is obtained by including the
electric field of the produced pairs.

In the Hamiltonian formalism spacetime is foliated by a series of spacelike
surfaces parametrized by a time variable.  For our purposes, it is
crucial that the surfaces pass smoothly through the horizon and are labelled
by coordinates which are free of singularities.  In principle the familiar
Kruskal coordinates could be used, as they satisfy these requirements, but
their rather complicated form leads to a cumbersome expression for the
effective action.  The advantage of the $L=1$, $R=r$ gauge in this context
is the simplification that is brought about by the vanishing of the terms
$\pi _R \dot{L} + \pi_R \dot{R}$.  In addition, even after quantization the
coordinate $r$ retains a clear physical meaning, as it can be inferred by
measuring the area of the constant $r$, $t$ 2-sphere.  In this gauge, then,
the model is simple enough, both conceptually and mathematically, to allow one
to obtain concrete results,
yet it plausibly contains the necessary ingredients
to provide for a realistic treatment of the radiation from a near extremal
hole.

\endpage

\refout

\figout

\end